\def\published#1{Pulished~#1\par}

\documentclass{ws-ijmpa}
\usepackage[super,compress]{cite}
\begin{document}

\markboth{K. Yi}
{review of $J/\psi\phi$}

\catchline{}{}{}{}{}

\title{
Experimental Review of  Structures in the $J/\psi\phi$ Mass Spectrum 
}

\author{Kai Yi
}

\address{Department of Physics and Astronomy, University of Iowa\\
Iowa City, Iowa 52242-1479 USA\\
yik@fnal.gov
}

\maketitle

\begin{history}
\received{16 April 2013}
\accepted{20 April 2013}
\published{23 July 2013}
\end{history}

\begin{abstract}

The discovery of numerous new charmonium-like structures since 2003  
have revitalized interest in exotic meson spectroscopy.
These structures do not fit easily into the conventional charmonium model,
and proposals like four-quark states, hybrids, and re-scattering effects 
have been suggested as explanations. 
 Since 2009, several new 
structures were reported in the $J/\psi\phi$ mass spectrum with the following  
characteristics:  they are the first ones reported  decaying into 
two heavy mesons which contain both a $c\bar{c}$ pair and a $s\bar{s}$ pair;
and their masses are well beyond the open charm pair threshold. 
Conventional $c\bar{c}$ states with a mass 
beyond the $J/\psi\phi$ threshold are not expected to decay into this channel 
and the width  is expected to be large, thus they are good candidates 
for exotic mesons.   My  focus in this 
article is to review the recent developments on the structures in the  $J/\psi\phi$ 
mass spectrum from CDF, Belle and LHCb.

\end{abstract}

\ccode{PACS numbers: 14.40.Gx, 13.25.Gv, 12.39.Mk}

\section{Introduction}	
\indent

The $q\bar{q}$ (meson) and $qqq$ (baryon) states predicted by the quark model 
have been well established experimentally~\cite{quarkmodel}. All ground-state mesons 
 have been 
observed since the discovery of the $B_c$ by CDF in 1998~\cite{Bc1,Bc2,Bc3,Bc4,Bc5}.

Hadrons  beyond  the $q\bar{q}$ and $qqq$ constructions are called  exotic 
states~\cite{jaff}.
The possibility of  $qq\bar{q}\bar{q}$ and $qqq\bar{q}\bar{q}$ states was
already suggested by Gell-Mann  at  the very birth of the quark model~\cite{gellmann}.  
With the development of QCD, 
the existence of extra exotic mesons such as glueballs (consisting purely of gluons) 
and hybrids (mixtures of valance gluons and quarks) was also 
proposed.\cite{glueball1,glueball2,glueball3,glueball4,glueball5,glueball6,glueball7,hybrid1,hybrid2,hybrid3,hybrid4,hybrid5,hybrid6}
There are a few candidates for exotic states in the light quark sector but 
none of them has been established convincingly.
There has also been  considerable theoretical work on exotic mesons  
in the heavy quark sector, especially
since the discovery of the $X(3872)$ by Belle in 
2003~\cite{x3872discoverybelle,x3872discoverycdf,x3872discoveryd0,x3872discoverybabar}.

\subsection{Recent Developments of Charmonium-like Structures}
\indent

In the charm sector, the charmonium potential model successfully explained 
and predicted the properties of charmonium states below the open charm 
threshold.\cite{charmpotential1,charmpotential2,charmpotential3} 
This model starts with a phenomenological potential which includes a 
Coulomb-like component and a  linear increasing component for quark confinement. 
It can be further extended to include spin-dependent terms and relativistic 
corrections. This model had been very successful for explaining the masses and widths 
for observed charmonium states---until the observation of the so-called 
charmonium-like states. The latter are superficially like a charmonium state 
but do not quite fit into the charmonium  model, for instance, 
their masses, widths, or decay rates  
deviate from charmonium expectations. Thus some of these structures are proposed
as candidates for various exotic mesons.

A possible experimental indication for the existence of exotic charmonium-like states
surfaced back in 1994 when
the CDF collaboration reported a $\psi(2S)$ production rate that was a 
factor of about 30 larger than the theoretical expectation~\cite{cdfpsiprime}.
It was soon suggested that an important part of this excess rate could be from contributions 
due to hybrid charmonium states.~\cite{closeCDF}
However, the color octet production mechanism is believed 
to be an  important factor contributing to the excess~\cite{octetcolorjpsi}. 
Even  though the feed-down from possible 
hybrids and other exotic states is not the dominant factor, 
it remains an open question whether exotics might make a substantial contribution
to inclusive charmonium production at hadron colliders.
It has also been proposed that the observed large branching fraction 
for non-charm $B$ decays  may indicate  a sizeable 
non-conventional charmonium production, such as  hybrid charmonium, in $B$ 
decays since the  dominant decay channel for conventional charmonium 
is open charm pair if the mass is above threshold~\cite{closeexclB}.

The discovery of the $X(3872)$ by Belle in 2003 opened  a new chapter for exotic 
mesons~\cite{x3872discoverybelle,x3872discoverycdf,x3872discoveryd0,x3872discoverybabar}, 
and subsequent discoveries~\cite{pdgrev} provide more candidates for exotic charmonium.  
The $X(3872)$ was discovered decaying into $J/\psi \pi^+\pi^-$, with a mass close 
to $DD^*$ threshold.  It was natural to consider it  a candidate 
for the missing  $D^3_2$ charmonium state.  
However, Belle searched further for the $D^3_2$ in its favored channel $J/\psi\gamma$, 
and found no signal~\cite{x3872discoverybelle}.
As pointed out in Belle's paper, the observed mass is higher than the theoretical 
prediction for the $D^3_2$, and the closeness of its 
mass to the $D\bar{D}^*$ threshold motivated speculation that this structure 
is some kind of exotic meson---perhaps a loosely bound `molecular' 
structure~\cite{x3872possibility1,x3872possibility2,x3872possibility3}. 
It is now ten years since the discovery of the $X (3872)$
and its nature is still not really resolved despite
a very large body of experimental work, including:
observation of a number of different decay modes~\cite{pdgrev},
precision mass measurements~\cite{cdfx3872mass,bellex3872mass}, studies of the 
$\pi^+\pi^-$ system in $J/\psi\pi^+\pi^-$ decays~\cite{cdfx3872pipi}, determination 
of its $J^{PC}$ as $1^{++}$~\cite{cdfjpc,babarjpc,Aaij:2013zoa}, and production 
characteristics~\cite{x3872discoveryd0,gerryrev,cmsx3872production,lhcbx3872production,Chatrchyan:2013cld}.

Many more charmonium-like states have been claimed since the  $X(3872)$, 
a recent review of the $X(3872)$ and the other 15 
observations can be found in Table 2 in Ref.~\cite{pdgrev}.
This spectroscopic bounty raises the question:
are there in fact too many of the so-called $X/Y/Z$  candidates, either as conventional 
or exotic charmonium?  Most of these reported structures need confirmation, 
and probably not all of them will survive further experimental tests, as pointed out,
for example, in Ref~\cite{toomany}.  Among those to be confirmed are structures 
like the $Z(4430)^+$ and the most recently reported   
$Z_c(3900)^+$~\cite{Ablikim:2013zna,Liu:2013dau,Xiao:2013iha}, 
which would be  {\it prima facie} smoking-guns for exotic states because  no 
charmonium state can, of course, be charged. 
Charmonium hybrids are also not an option for these observations due to their 
charges and they are likely to be a four-quark state.  
However, they need to be first confirmed as resonance states and 
their neutral partners need to be established before settling on a 
four-quark interpretation.

The structures reported in the $J/\psi\phi$ mass spectrum 
are also striking. 
Their widths are relatively narrow and their masses are well beyond the open charm pair 
threshold, to which conventional charmonium of such high mass should overwhelmingly 
prefer to decay into~\cite{charmpotential1,charmpotential2,charmpotential3}, and thus these also strongly imply
an exotic nature.
They are also the first structures reported 
which decay into two heavy mesons that contain both a $c\bar{c}$ 
pair and an  $s\bar{s}$ pair.
Interestingly, the $J/\psi\phi$ system has positive C parity---since
the two decay daughters both have $J^{PC}=1^{--}$---which keeps
the door open for this state to be  manifestly 
exotic, i.e. $J^{PC}=1^{-+}$ is forbidden for conventional $c\bar{c}$. A recent lattice 
calculation for a $1^{-+}$ charmed hybrid predicts a mass of $4.30\pm0.05$ GeV, 
which also places these reported structures  in the right mass 
range~\cite{myccbaroneminuspluslatticeqcd}.
A  $1^{-+}$ charmonium-like state would therefore also provide a smoking-gun 
for an exotic meson.

In a word, we are in exciting times for heavy-quark exotic spectroscopy, 
many exotic candidates  have been observed but none 
of them can be declared as conclusive yet.  Besides the observed 
charmonium-like structures, there are also bottomonium-like structures 
observed~\cite{pdgrev}, and all these together provide 
us rich opportunity to understand   exotic  mesons.
In this article I review recent developments 
on the structures in the  $J/\psi\phi$ mass spectrum from CDF, Belle and LHCb.

\subsection{The Currently Reported Structures in the  $J/\psi\phi$ Spectrum}
\indent

\begin{table}[ph]
\tbl{Structures reported in the $J/\psi\phi$ mass spectrum.
}
{\begin{tabular}{@{}cccc@{}} \toprule
Production Process  & Experiment & Mass (MeV) & Width (MeV) \\
$B^+\rightarrow J/\psi\phi K^+$ & CDF & $4143.4^{+2.9}_{-3.0} (\mathrm{stat})\pm 0.6(\mathrm{syst})$  & $15.3^{+10.4}_{-6.1}(\mathrm{stat})\pm 2.5 (\mathrm{syst})$  \\
%$B^+\rightarrow J/\psi\phi K^+$  & CMS & $4148.2{\pm2.0} (\mathrm{stat})\pm 4.6(\mathrm{syst})$  & ---  \\
$B^+\rightarrow J/\psi\phi K^+$ & CDF & $4274.4^{+8.4}_{-6.7}(\mathrm{stat})\pm1.9(\mathrm{syst})$  &  $32.3^{+21.9}_{-15.3}(\mathrm{stat})\pm7.6(\mathrm{syst})$  \\
%$B^+\rightarrow J/\psi\phi K^+$  & CMS & $4316.7{\pm3.0}(\mathrm{stat})\pm7.3(\mathrm{syst})$  &  ---  \\
$e^+e^-\rightarrow \gamma\gamma e^+e^-$ & Belle & $4350.6^{+4.6}_{-5.1}(\mathrm{stat})\pm0.7(\mathrm{syst})$  & $13^{+18}_{-9}(\mathrm{stat})\pm4(\mathrm{syst})$  \\  \botrule
\end{tabular} \label{jpsiphistructures}}
\end{table}

The CDF collaboration reported the first evidence of a new structure near the 
$J/\psi\phi$ threshold, the $Y(4140)$, in 2009 through 
exclusive $B^+ \rightarrow J/\psi\phi K^+$ decays~\cite{cdfevidence}.
In a subsequent report  CDF placed the significance of 
the structure as exceeding 5$\sigma$~\cite{cdfobservation}. In addition to this
structure, CDF also reported  evidence for a second one 
around 4.28 GeV, 
which is curiously about one pion mass 
above the reported $Y(4140)$. 
In 2010 the Belle collaboration reported evidence for a structure around 4.35 GeV 
in a search through two-photon processes~\cite{belleevidence}. 
The situation in $J/\psi \phi$ structures became murkier in 2011 when LHCb 
reported a null search for the $Y(4140)$ in conflict with CDF's report~\cite{lhcbjpsiphik}.
However, in 2012, the CMS collaboration reported preliminary observations of two structures 
in the $J/\psi\phi$ spectrum from  exclusive $B^+\rightarrow J/\psi \phi K^+$ 
decays. The lower mass structure from CMS provides the first confirmation of the existence 
of the $Y(4140)$. 
CMS's second structure is close to CDF's $Y(4274)$, but the difference in mass is large enough,
given the quoted errors, such that it does not obviously support equating the two structures.
The properties of these structures
are summarized in Table \ref{jpsiphistructures}.

\subsection{Production of Possible $J/\psi\phi$ Structures}
\indent

The $J/\psi\phi$ system consists of two $1^{--}$ states, $J/\psi$ and $\phi$, and therefore
depending on the orbital angular momentum configuration  the system 
can have the following $J^{PC}$: $0^{++}$, $1^{++}$, and  $2^{++}$ for $s$-wave 
coupling; and $0^{-+}$, $1^{-+}$,   $2^{-+}$, and  $3^{-+}$ for $p$-wave. 
As noted above, the  $J/\psi\phi$ system
always carries positive C-parity, and therefore it cannot be 
produced through processes that carry negative C-parity, such as 
Initial State Radiation (ISR) in an $e^+e^-$ collider where either electron or positron radiate 
a photon before the  electron-positron annihilation~\cite{isr1,isr2,isr3,isr4}.    
The $J\psi\phi$ system is mainly accessible in the following processes: exclusive $B$ 
decays ($b\rightarrow c\bar{c}s$ transition),
two-photon process, and prompt production in $p\bar{p}/pp$ interactions.

{ \bf Exclusive $B\rightarrow J/\psi \phi K$ decays}: 
For the exclusive $B\rightarrow J/\psi \phi K$ decay, 
the creation of an $s\bar{s}$ from the vacuum is needed,  
so the branching fraction is suppressed ($OZI$ rule).
Supposing that there is no structure in any two-body 
sub-system ($J/\psi \phi$, $J/\psi K^+$, $\phi K^+$),
the decay of  $B\rightarrow J/\psi \phi K$ will be simply kinematically constrained so 
that the $J/\psi \phi$ mass spectrum will be just phase space.
Other sorts of production, such as possible hybrid charmonium~\cite{closeexclB} or tetra-quark 
states,~\cite{riccardo} appear as  structures on top of phase space. 
A reasonable production rate of exotic states can be prominent 
over a small phase space background, thus making this decay a promising channel for 
exotic states searches. 
The long $B$ lifetime provides an additional experimental handle 
to obtain a clean $B$ signal. Compared to inclusive production, which will be 
discussed below, the narrow $B$ 
signal  and its sideband can provide constraints 
on possible reflections from partially reconstructed  or 
mis-reconstructions of other $B$ hadrons. The latter is because the 
partial reconstruction  or mis-reconstruction of other $B$ hadrons 
are expected to extend well into  $B$ sideband.

{ \bf Two-photon processes}: In  $e^+e^-$ collisions
the electron and positron can each radiate a 
photon which interact with each other to produce new particles. 
The mass spectrum of  
$J/\psi\phi$  from $\gamma\gamma$ interactions can be used to 
search for the states otherwise inaccessible to  $e^+e^-$ annihilation. 
Normally, these events are selected by requiring 
a very low transverse momentum of the system being searched for, 
which provides a very clean signal. However, producing a state of $3^{-+}$ 
as well as states with $J=1$ 
states are forbidden in this process by Yang's Theorem~\cite{yang}.

{\bf Prompt production from $p\bar{p}/pp$ interactions}:
All $J^{PC}$ states are allowed in this process.  
Depending on the prompt production rate and the width of the possible 
structures, inclusive searches in prompt production can be important 
in $p\bar{p}/pp$ interactions---e.g., the first confirmation of the $X(3872)$ 
by CDF~\cite{x3872discoverycdf}.  
However, it is in general difficult to deal with large prompt backgrounds, 
and the lack of constraints can make it difficult to understand possible reflections.

\section{Observation of $B\rightarrow J/\psi \phi K$}
\indent

\begin{figure}[pb]
{\psfig{file=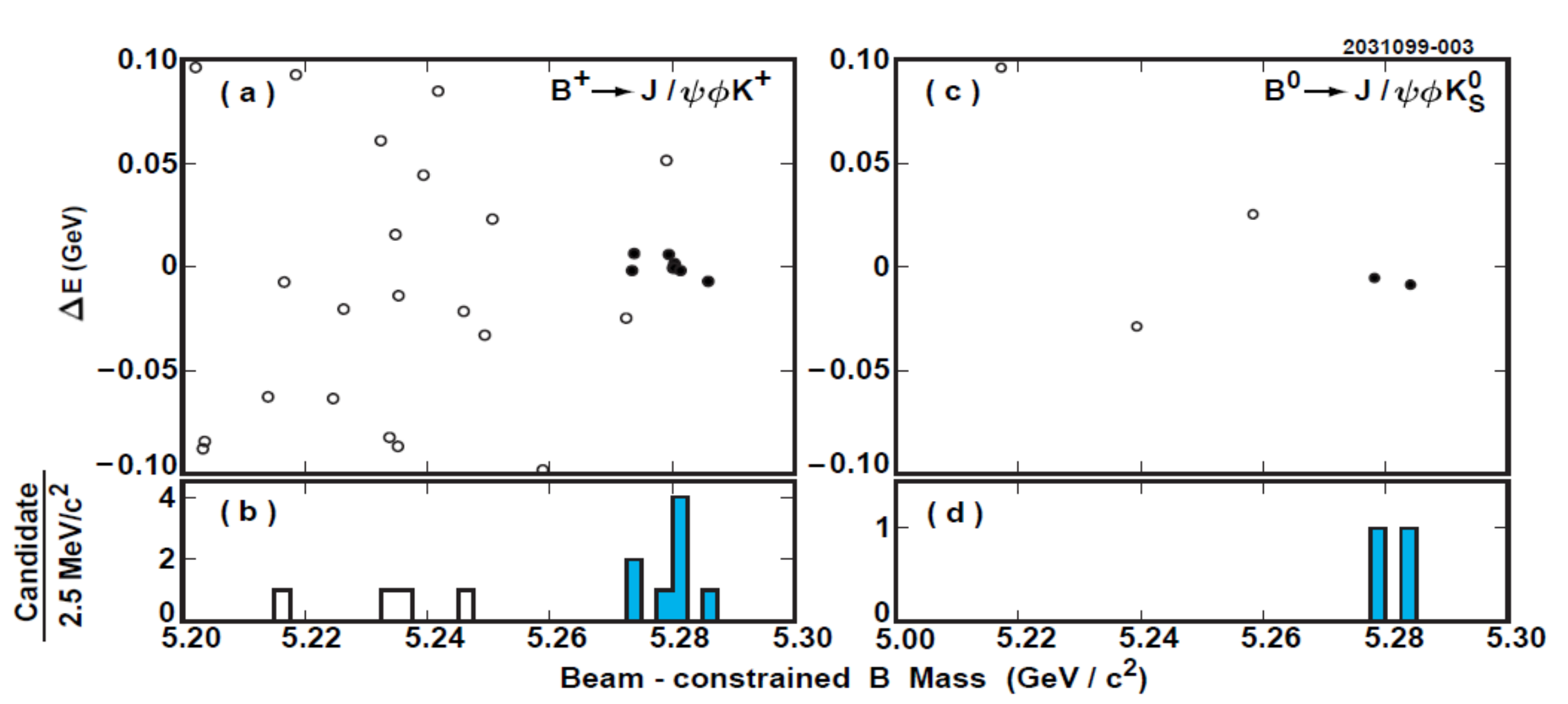,width=0.69\textwidth}}
{\psfig{file=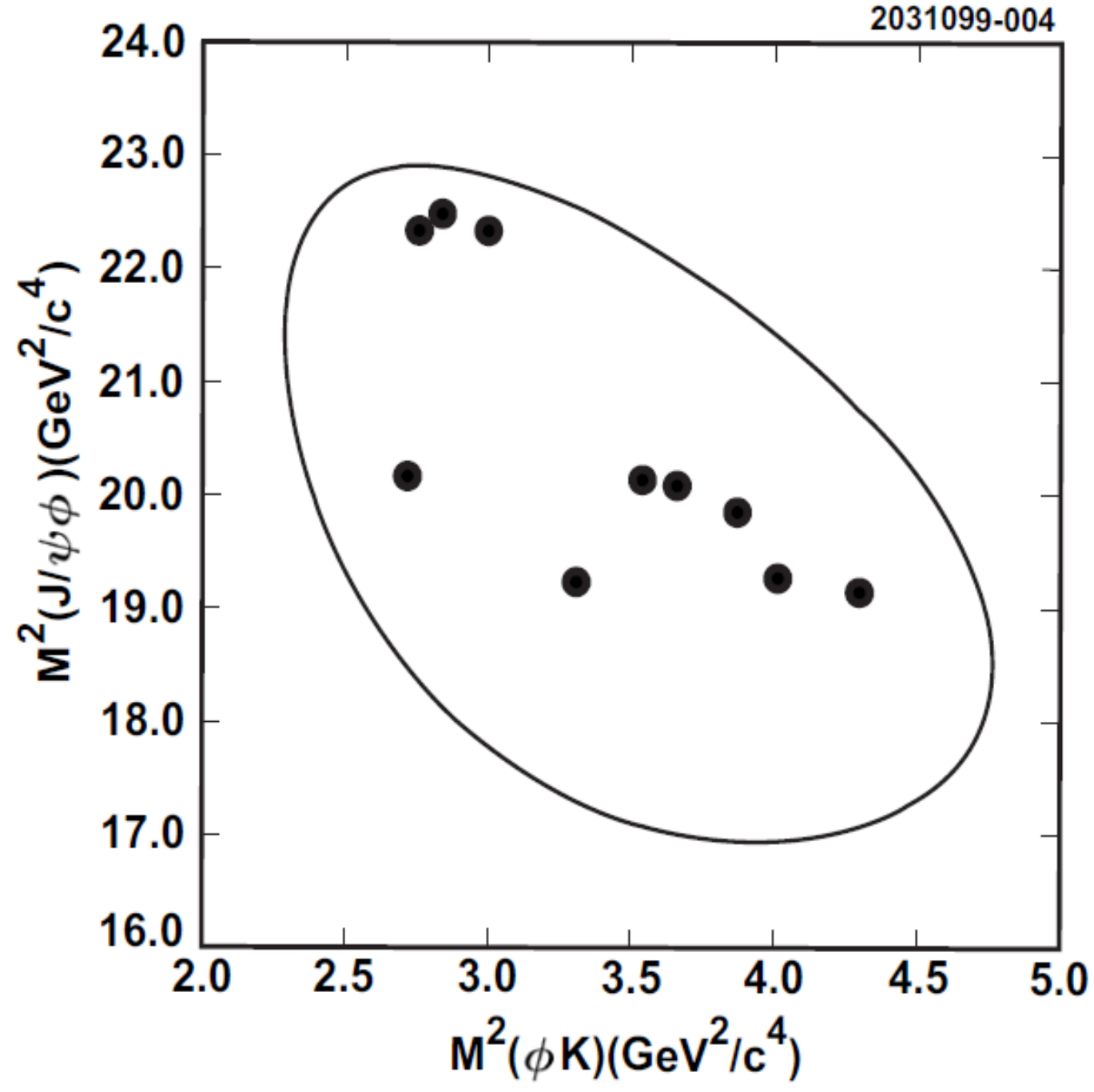,width=0.30\textwidth}}
\vspace*{8pt}
\caption{
The $\Delta E$  vs $M(B)$ distribution for (a) $B^+\rightarrow J/\psi\phi K^+$ 
and (c) $B^0 \rightarrow J/\psi \phi K^0_S$ candidates in data collected by the CLEO 
experiment. $B$-signal candidates are shown by filled circles,
and background by the open ones. 
Below the scatter plots are
the $M(B)$ distribution for (b) $B^+\rightarrow J/\psi\phi K^+$ 
and (d) $B^0 \rightarrow J/\psi \phi K^0_S$ candidates 
in the selected $\Delta E$ range. The signal candidates are shown as  
the shaded parts of the histograms.
On the far right is
the Dalitz plot for the 10 $B\rightarrow J/\psi\phi K$ 
candidates, with the
solid line marking the kinematic boundary.
\label{f1cleof3}}
\end{figure}

The first reports of 
exclusive $B^+\rightarrow J/\psi \phi K^+$  and $B^0\rightarrow J/\psi \phi K^0_S$
signals were from the CLEO collaboration in 1999 with 9.1 fb$^{-1}$ of 
 $e^+e^-$ data~\cite{cleobobservation}. 
Figure~\ref{f1cleof3} (left and middle) shows the $\Delta E$ vs $M(B)$ distributions
reported by CLEO, where  $\Delta E$ is defined 
as $E(J/\psi)+E(\phi)+E(K)- E_{beam}$ and  $M(B)$ is 
$\sqrt{E^2_{beam}-p^2(B)}$.
The Dalitz plot for the selected $B$ signal 
events is shown in Fig.~\ref{f1cleof3} (right). There was no discussion on possible structures 
in the $J/\psi\phi$ mass spectrum due to low statistics even though 
in hindsight it seems that those 
10 events do concentrate in two clusters.  
Furthermore,  as pointed out in CLEO's report, their
reconstruction efficiency  was close to zero near the edge of  phase space, 
and thus they were  not sensitive to possible structures 
in the low $J/\psi\phi$ mass region below the $DD^{**}$ threshold.

\begin{figure}[pb]
{\psfig{file=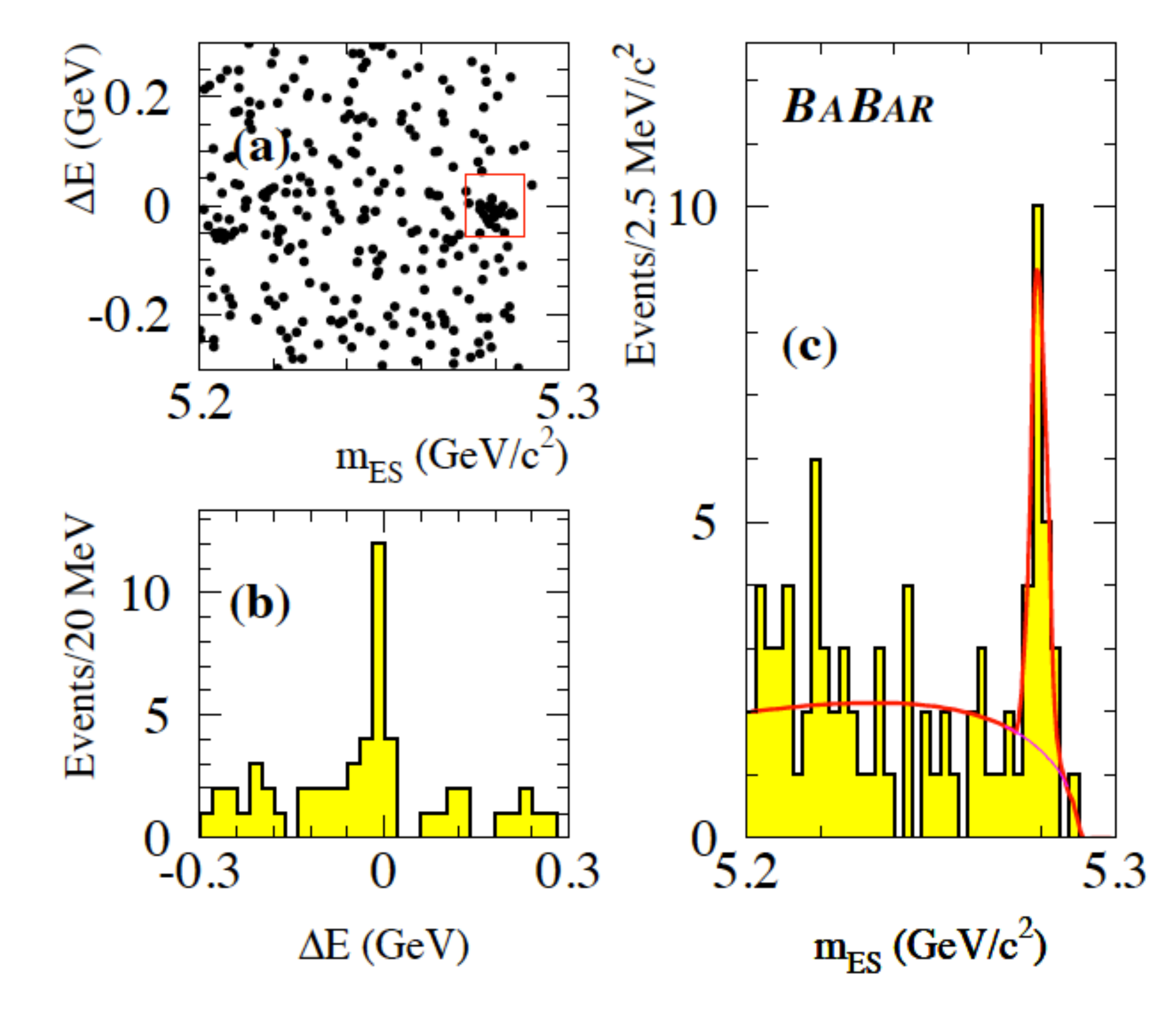,width=0.49\textwidth}}
{\psfig{file=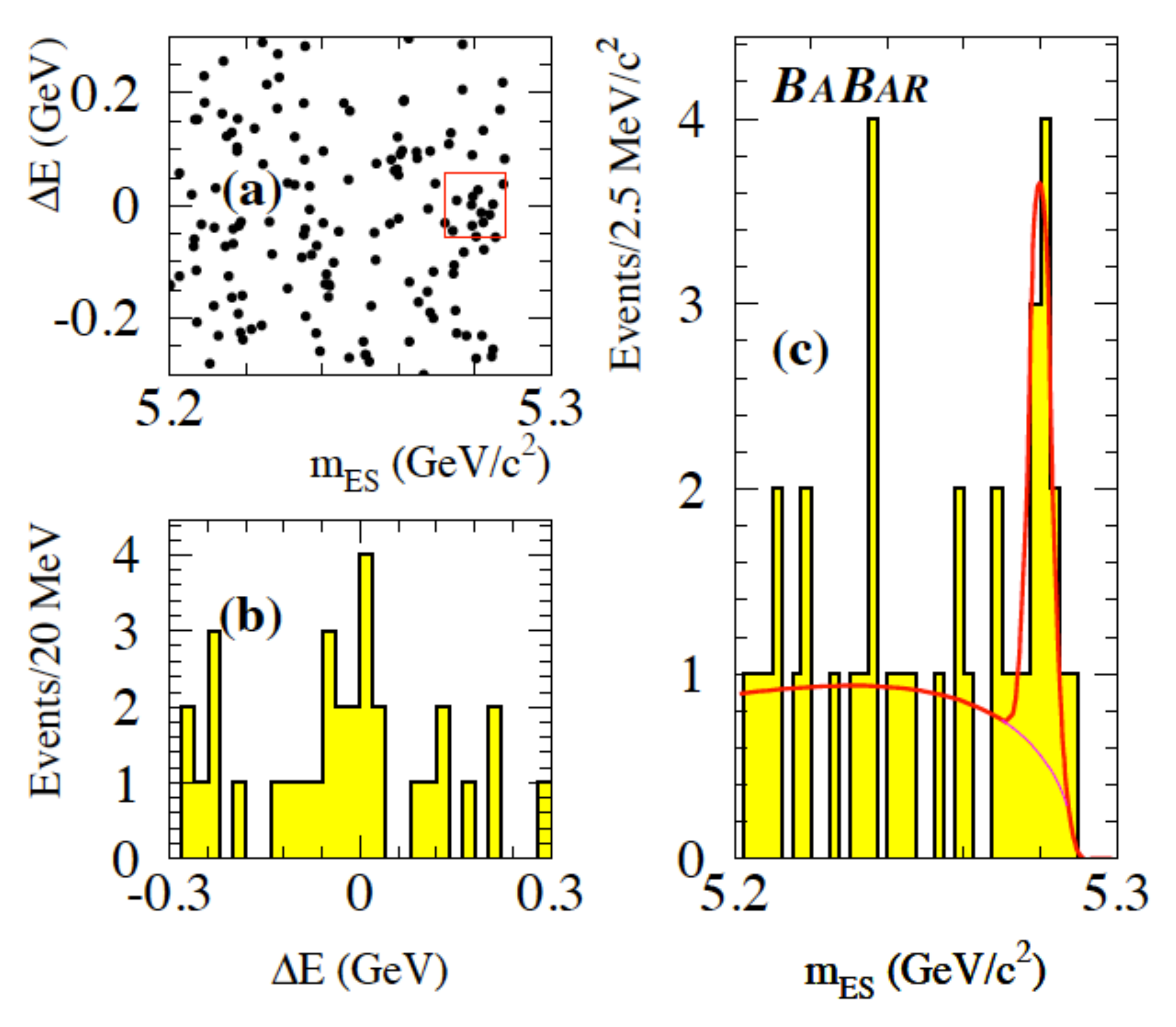,width=0.49\textwidth}}
\vspace*{8pt}
\caption{
The $\Delta E$ and $m_{ES}$ distributions for $B^+ \rightarrow J/\psi \phi K^+$ (left set) 
and $B^0 \rightarrow J/\psi \phi K^0_S$ (right set) from BaBar.  
For each respective mode,
the $\Delta E$ vs $m_{ES}$ event distribution is shown in (a) with a small 
rectangle as the signal box. The $\Delta E$ projection of the $m_{ES}$ signal 
region selection is shown in (b). The $m_{ES}$ projection of the $\Delta E$ signal 
region selection is shown in (c). 
The red solid lines in (c) of both left and right plots are fits to the data with background 
and signal.
}
\label{f3babarf1}
\end{figure}

The BaBar collaboration reported more significant $B^+\rightarrow J/\psi\phi K^+$ (23 events)
and $B^0 \rightarrow J/\psi \phi K^0_S$ (13 events) signals with 50.9 fb$^{-1}$ of 
data~\cite{babarbobservation}. 
The $\Delta E$ 
vs $m_{ES}$ [{i.e.} $M(B)$] distribution and the projections for both $B^+$ and $B^0$ are shown 
in Fig.~\ref{f3babarf1}.  There was again no report 
on the examination of the $J/\psi\phi$ mass spectrum due to the low statistics.

Four other experiments have also observed  significant $B \rightarrow J/\psi \phi K$ signals,
but these reports are intimately involved in the matter of positive evidence 
for structures in the $J/\psi \phi$ mass distribution and are the topic of the next section.

\section{Evidence of Structures in the $J/\psi\phi$ Mass Spectrum}
\indent

In this section we discuss the evidence of structures in the 
 $J/\psi\phi$ mass spectrum from exclusive $B$ decays, 
and from two-photon process, based on reports by CDF, Belle, and 
LHCb---but first we briefly review the detectors involved.

\subsection{The CDF, Belle and LHCb  Detectors}
\indent

\subsubsection{The CDF Detector}
\indent

CDF collected data at the
Tevatron, a $p\bar{p}$ collider at a center-of-mass energy of 2 TeV, 
with an instantaneous luminosity ranging from the initial $10^{31}$ 
to $4.3\times 10^{32}$  cm$^{-2}$ sec$^{-1}$ 
at the end of its program. 
The  cross section for $B$ hadron production at the Tevatron is very large,
around 20 $\mu b$ in the kinematic region of interest. 
CDF II was a general purpose solenoidal detector which combines precision 
charged particle tracking with fast projective calorimetry and fine grained muon 
detection~\cite{cdfdetecto1,cdfdetecto2}. 
It was nearly cylindrically 
symmetric with respect to the beamline, and forward-backward symmetric with 
respect to the nominal interaction point. The tracking system was in a 1.4 T 
axial magnetic field provided by a superconducting solenoid.  
Charged particles created in the $p\bar{p}$ collision
had their momentum and charge measured by the tracking system. 
The sub-tracking system, the Central-Outer-Tracker (COT), measured $dE/dx$ for charged 
particles which could be used for hadron particle identification. Right outside the tracking 
system, there was a Time-of-Flight detector which was of further aid 
in hadron particle identification, especially  for studying  $J/\psi\phi$ system. 
The muon chamber system was the outermost part of CDF, and  was used to trigger and 
identify muons. 
A crucial component of the  system for $B$ studies was the Silicon Vertex Detector, 
which made it possible to reconstruct the $B$ signal with excellent mass resolution and 
isolate the $B$ signal from large prompt background using its excellent vertex resolution.
The typical mass resolution for $B$ hadrons was 5-10 MeV, 
and the typical 
vertex resolution was about 20-30 $\mu m$.

The trigger system plays a crucial role in a hadron collider experiments because 
the interaction rate is so high that each collision can not be read out and recorded.
The data must be filtered to obtain interesting events. Storage capacity  
limited the trigger rate to about 75 Hz.
The online physics event selection at CDF was achieved by a three-layer trigger.
The CDF  trigger relevant to  $J/\psi\rightarrow \mu^+\mu^-$
required two  oppositely charged online muon tracks with $p_T>1.5$ or $p_T>2.0$  GeV, 
depending on the part of the muon system involved, 
and the dimuon mass was required to be in the range from 2.7 to 4.0 GeV. 
This trigger was dynamically pre-scaled due to its high rate.

Due to the transverse momentum requirement in the trigger system and the 
centrality of the detector, the $B\rightarrow J/\psi\phi K$ hadrons were
highly boosted (the typical $B$ transverse momentum was above 4 GeV), 
thus providing high efficiency 
for $B$ events even near the edge of their decay phase space.  
This is an  important characteristic of a central hadron-collider detector and a significant 
factor in searching for structures in the near-threshold region in the $J/\psi\phi$ 
mass spectrum.

\subsubsection{The Belle Detector}
\indent

The Belle detector is installed at  an asymmetric $e^+ e^-$ collider 
operating at the $\Upsilon(4S)$ resonance, the KEK-B-factory~\cite{belledetector}.
It is a large-solid-angle magnetic spectrometer that 
consists of a silicon vertex detector (SVD), a 50-layer central drift 
chamber (CDC), an array of aerogel threshold Cherenkov counters (ACC), 
a barrel-like arrangement of time-of-flight scintillation counters (TOF), 
and an electromagnetic calorimeter comprised of CsI(Tl) crystals (ECL)
located inside a superconducting solenoid coil that provides a 1.5 T 
magnetic field. An iron flux-return located outside of the coil is 
instrumented to detect $K^0_L$ mesons and to identify muons (KLM).

The trigger system at Belle is almost 100\% efficient for $B$ mesons.
The Belle detector has excellent mass resolution, quite similar to CDF's. 
Belle's particle identification is intrinsically very powerful for the low transverse momentum 
tracks from $B$ decays.

\subsubsection{The LHCb Detector}
\indent

The LHCb detector is running at the Large Hadron Collider (LHC), which 
provides $pp$ collisions at a center-of-mass energy up to 8 
TeV (so far)~\cite{lhcbdetector}.
It is a single-arm forward spectrometer covering the pseudorapidity 
range $2 < \eta < 5$, which  includes a high precision silicon tracking system and 
straw drift-tubes. The dipole magnet provides 1.4 T  
magnetic field.
The combined tracking system has momentum resolution $\Delta p/p$ that varies 
from 0.4\% at 5 GeV and  to 0.6\%  at 100 GeV, and an impact parameter (IP) 
resolution of 20 $\mu m$ for tracks with high transverse momentum. 
Charged hadrons are identified using two ring-imaging Cherenkov 
detectors. Photon, electron, and hadron candidates are identified by a calorimeter system
consisting of scintillating-pad and pre-shower detectors, an electromagnetic calorimeter
(ECAL), and a hadronic calorimeter (HCAL).
Muons are identified by a muon system (MUON) composed of alternating layers of 
iron and multi-wire proportional chambers.
The MUON, ECAL, and HCAL provide the capability for first-level hardware triggering.
The single and dimuon hardware triggers provide good efficiency for $B$ hadrons. 
A single track trigger and dimuon trigger are used in the analysis discussed here. 
At the final stage, they either require a $J/\psi \rightarrow \mu^+\mu^-$ 
with $p_T>1.5$ GeV or a muon-track pair with significant IP.

The LHCb detector has excellent mass resolution and excellent hadron particle 
identification ability. The mass resolutions of $B^+\rightarrow J/\psi \phi K^+$ 
for both CDF and LHCb are both between 5 and 6 MeV, while the Belle experiment 
obtains better resolution due to the beam energy constraint.
The LHCb experiment handles a very high data rate with an open 
trigger which enables them to collect a very broad spectrum of hadronic $B$ decays.
However, as a result the LHCb experiment runs at a relatively  low instantaneous luminosity 
compared to the  ATLAS and CMS experiments due to limitations on the trigger bandwidth.
Different from most other detectors, which measure the transverse 
momentum of tracks, LHCb mainly measures 
a track's longitudinal momentum.

\subsection{Structures in the $J/\psi\phi$ Mass Spectrum: CDF }
\indent

\begin{figure}[pb]
\centerline{\psfig{file=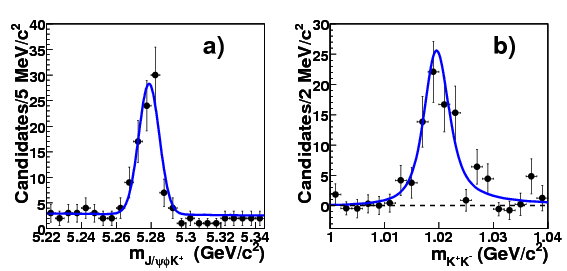,width=0.9\textwidth}}
\vspace*{8pt}
\caption{
(a) The mass distribution of $J/\psi\phi K^+$ from CDF with 2.7 fb$^{-1}$ of data; the solid line
is a fit to the data with a Gaussian signal function and flat
background function. (b) The $B^+$ sideband-subtracted mass
distribution of $K^+K^-$ without the  mass window requirement, where
the solid curve is a P-wave relativistic Breit-Wigner
fit to the data.
\label{f5cdff1}}
\end{figure}

\begin{figure}[pb]
\centerline{\psfig{file=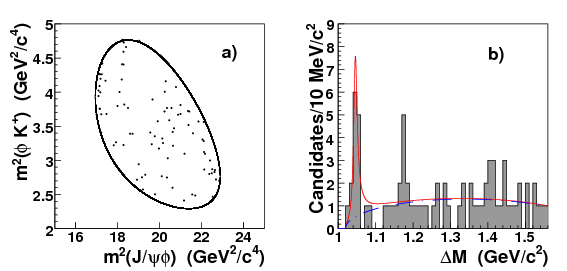,width=0.9\textwidth}}
\vspace*{8pt}
\caption{
(a) The Dalitz plot of $m^2(\phi K^+)$ versus $m^2(J/\psi\phi)$
in the $B^+$ mass window from CDF with 2.7 fb$^{-1}$ of data.
The boundary shows the kinematically allowed region.
(b) The mass difference, $\Delta M$, between $\mu^+\mu^-K^+K^-$ and $\mu^+\mu^-$, 
in the $B$ mass window from CDF. The dash-dotted
curve is the background contribution and the red-solid
curve is the result of the unbinned signal-plus-background  fit
assuming a single state.
}
\label{f6cdff2}
\end{figure}

The CDF collaboration reported  evidence of a structure near the $J/\psi\phi$ 
threshold in exclusive $B^+\rightarrow J/\psi \phi K^+$ using 2.7 fb$^{-1}$
of data in 2009~\cite{cdfevidence}. 
The CDF analysis first reconstructed a $J/\psi\rightarrow \mu^+\mu^-$ candidate 
by forming the dimuon tracks into a common vertex, and then 
required the $B$ meson to be 500 $\mu$m away from the 
primary vertex in the transverse plane, which reduces the background by a factor of 
200 and keeps the signal with a efficiency of 60\%. The track $dE/dx$ and Time-of-flight 
information is summarized in a log-likelihood ratio, which reflects how well a candidate 
track can be positively identified as a kaon relative to a pion.  The requirement of 
minimum log-likelihood ratio of 0.2 for all three kaon candidates reduces the non-B 
background by a factor of 100 with an efficiency of about 40\%. 
The  mass distribution of $J/\psi\phi K^+$ after all selections 
is shown in Fig.~\ref{f5cdff1} a) 
with $75 \pm 10$ $B$ signals, 
and the $B$ signal purity is about 80\% with a mass resolution 
of 6~MeV.
Figure~\ref{f5cdff1} b) shows the  $B^+$ sideband-subtracted mass
distribution of $K^+K^-$ without the  $\phi$ mass-window requirement. 
A fit was done to the  $K^+K^-$ distribution using a Breit-Wigner function 
only and it returns a 
mass and width compatible with the $\phi$ meson, which shows that $J/\psi \phi K^+$   
dominates the observed $B$ signal.
 Figure~\ref{f6cdff2} a) shows the Dalitz plot 
for those events in the $B$ mass window. There is a cluster around the 
$m^2(J/\psi\phi)$ of 17 GeV$^2$, and there may be another cluster around 18 GeV$^2$.

\begin{figure}[pb]
{\psfig{file=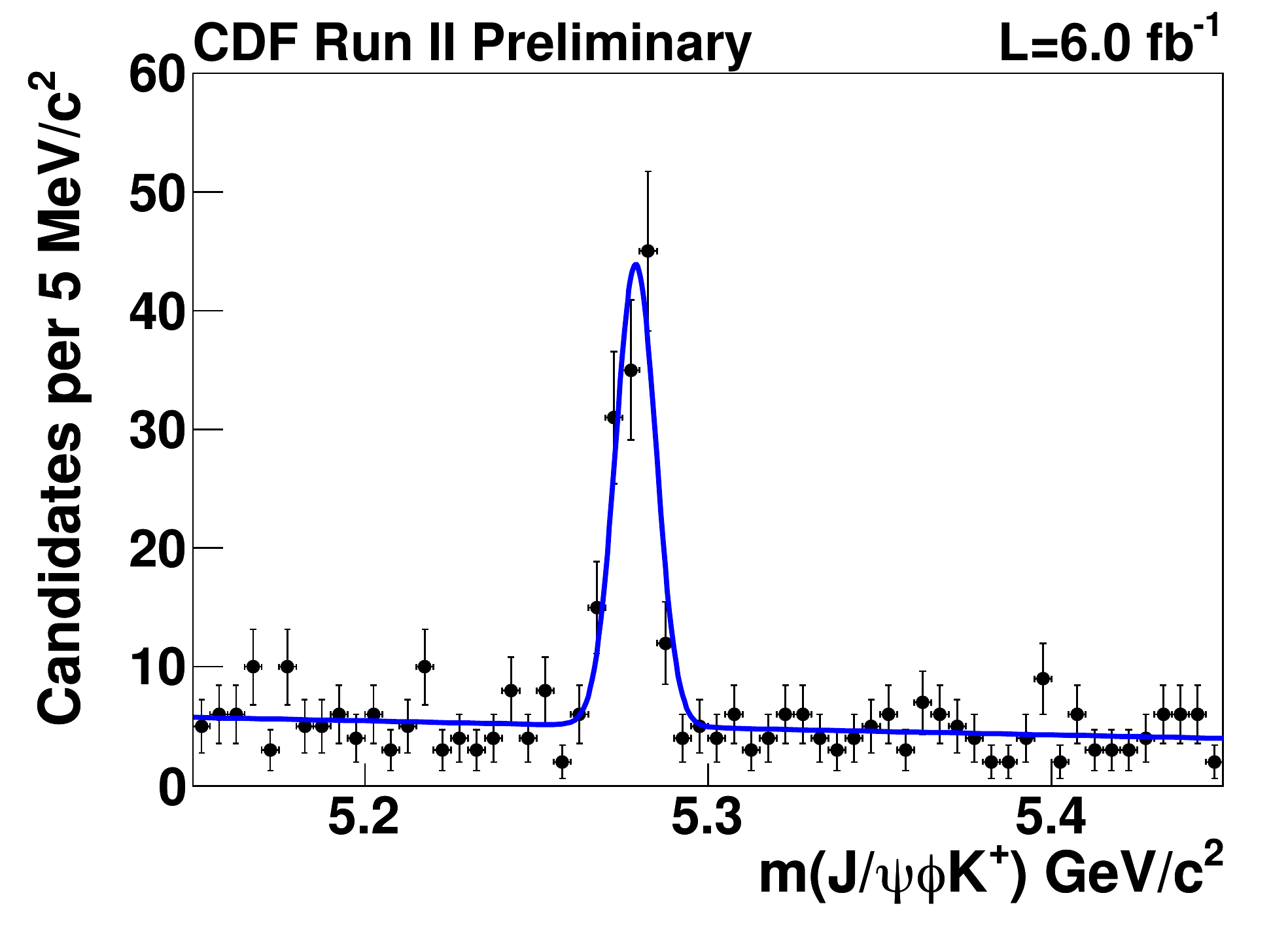,width=0.48\textwidth}}
{\psfig{file=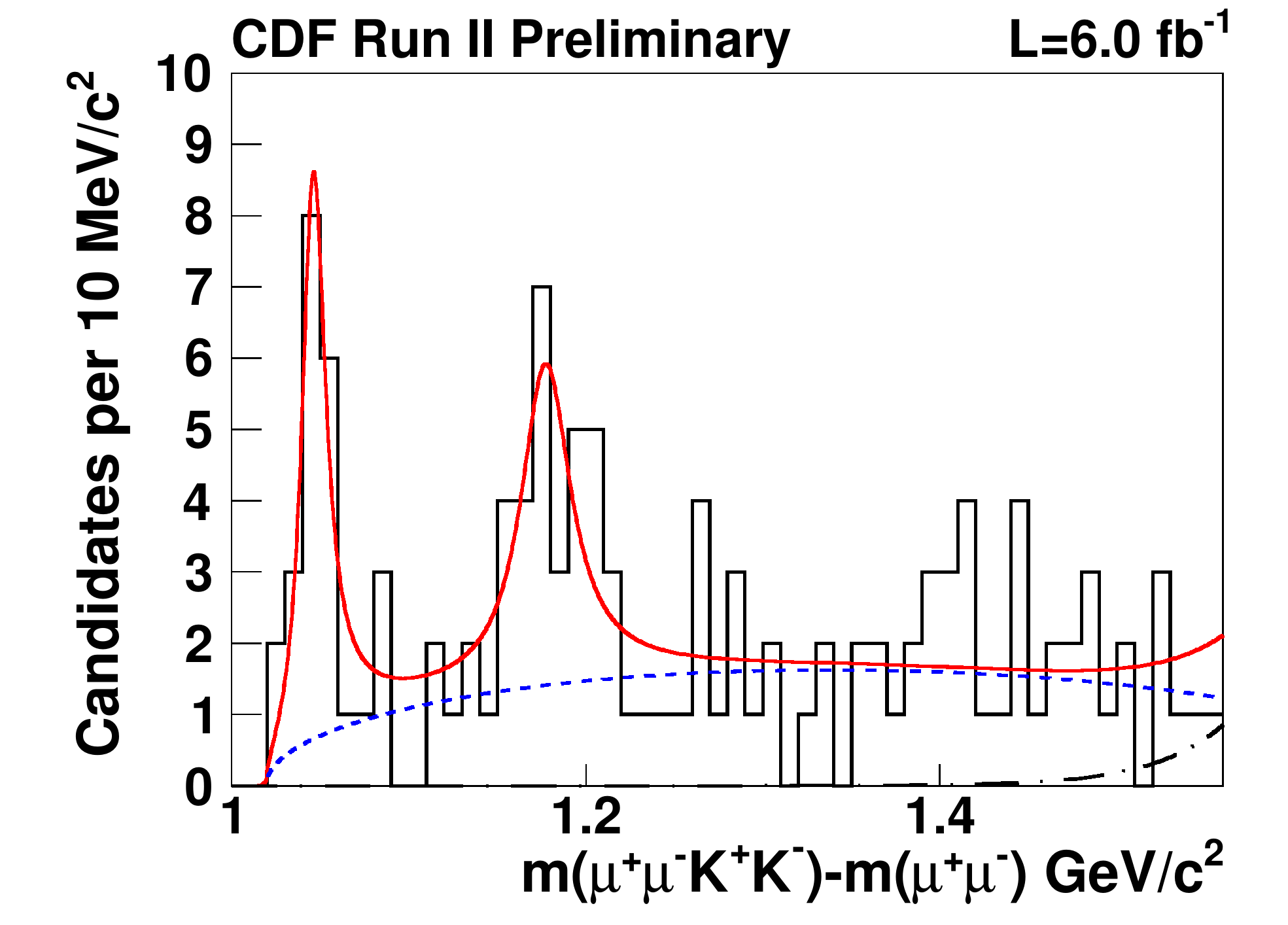,width=0.48\textwidth}}
\vspace*{8pt}
\caption{
(a) The mass distribution of $J/\psi\phi K^+$ from CDF with 6.0 fb$^{-1}$ of data; the solid line
is a fit to the data with a Gaussian signal function and flat
background. 
(b) The mass difference, $\Delta M$, between $\mu^+\mu^-K^+K^-$ and $\mu^+\mu^-$,
in the $B$ mass window from CDF with 6.0 fb$^{-1}$ of data. The dashed
curve is the background contribution, the dashed-dotted curve is the contribution 
from the contamination from $B_s\rightarrow \psi(2S) \phi$ and the red-solid
curve is the total unbinned fit incorporating two states.
\label{f7cdff3}}
\end{figure}

Figure~\ref{f6cdff2} b) shows the Dalitz projection onto $m(J/\psi\phi)$ 
but expressed as the mass difference $\Delta M=m(\mu^+\mu^-K^+K^-)-m(\mu^+\mu^-)$. 
A prominent structure appeared near the $J/\psi\phi$ threshold with a 
mass of $4143.0 \pm 2.9 (\mathrm{stat})\pm 1.2(\mathrm{syst})$ MeV 
and a width of $11.7^{+8.3}_{-5.0}(\mathrm{stat})\pm 3.7 (\mathrm{syst})$ MeV, 
which,
following the pattern of other $X$ and $Y$ structures, was labeled the $Y(4140)$.
With increased statistics over the original report, i.e. 6.0 fb$^{-1}$ of data and $115\pm12$~$B$ 
events,
the statistical significance of the $Y(4140)$ 
exceeded 5$\sigma$ with 19$\pm$6 signal events, assuming a relativistic BW 
for the signal and three--body phase space for the background as before.
The $J/\psi\phi K^+$ invariant mass 
distribution and $\Delta M$ for the  6.0 fb$^{-1}$  are shown in Fig.~\ref{f7cdff3}. 
The mass and width of this structure are updated  as 
$4143.4^{+2.9}_{-3.0} (\mathrm{stat})\pm 0.6(\mathrm{syst})$ MeV
and $15.3^{+10.4}_{-6.1}(\mathrm{stat})\pm 2.5 (\mathrm{syst})$ MeV. 
The relative branching fraction of $Y(4140)$ over inclusive 
$B^+\rightarrow J/\psi\phi K^+$ decays 
$\mathcal{B}_{rel}=\mathcal{B}(B^+\rightarrow Y(4140) K^+) \times \mathcal{B}(Y(4140)\rightarrow J/\psi\,\phi)$/$\mathcal{B}(B^+\rightarrow J/\psi\,\phi K^+)$ including systematic uncertainties is 
measured as 
$0.149\pm 0.039 (\mathrm{stat})\pm 0.024 (\mathrm{syst})$~\cite{cdfobservation}.

In addition, CDF also reported 
evidence for another structure in this larger data set around $\Delta M=1.18$ GeV 
with a significance of 3.1$\sigma$ and a signal yield of $22\pm8$.  
The fitted mass difference and width are $1177.7^{+8.4}_{-6.7}$ MeV and $32.3^{+21.9}_{-15.3}$ MeV.
Including systematics, 
the mass of this structure is $4274.4^{+8.4}_{-6.7}(\mathrm{stat})\pm1.9(\mathrm{syst})$  MeV 
by adding the $J/\psi$ mass and including systematics, and the width 
is measured as $32.3^{+21.9}_{-15.3}(\mathrm{stat}) \pm 7.6 (\mathrm{syst}) $ MeV

\subsection{Structures in the $J/\psi\phi$ Mass Spectrum: Belle }
\indent

In 2009, the Belle collaboration promptly searched for the $Y(4140)$ structure through the same 
exclusive $B^+ \rightarrow J/\psi \phi K^+$ decays 
using 825 fb$^{-1}$ of $\Upsilon(4S)$ data~\cite{bellejpsiphik1,bellejpsiphik2}.
Belle found $7.5^{+4.9}_{-4.4}$ signal candidates from $325\pm21$~$B$ events, 
corresponding to a significance of 1.9$\sigma$,
by imposing the $Y(4140)$ parameters from CDF in their fit.  
This clearly can not count as a confirmation.
However, the efficiency near the $J/\psi \phi$ threshold 
is low 
and the Belle collaboration 
stated that they can neither confirm nor exclude the existence of the $Y(4140)$
with their current data~\cite{bellejpsiphik2}.

\begin{figure}[pb]
{\psfig{file=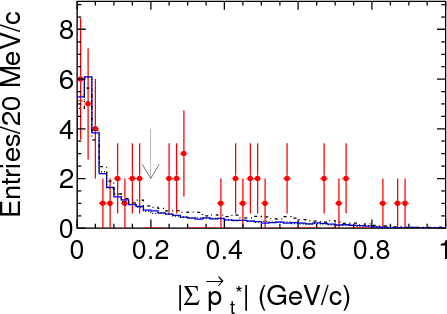,width=0.49\textwidth}}
{\psfig{file=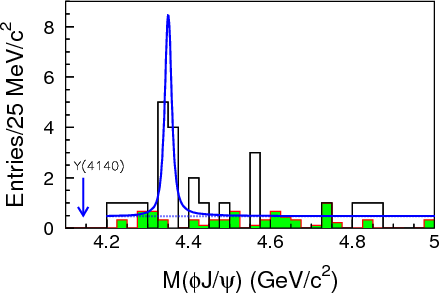,width=0.49\textwidth}}
\vspace*{8pt}
\caption{
Left: the magnitude of the vector sum of the $J/\psi\phi$ transverse momentum with respect 
to the beam direction in the $e^+e^-$ center-of-mass frame for 825 fb$^{-1}$ of Belle data,
which are represented as points with error bars.  MC simulation 
for $\gamma\gamma\rightarrow J/\psi\phi$ with a mass fixed at 4.20, 4.35, and 4.50 GeV 
are represented as  dot-dashed, solid, and dotted histograms.
The arrow shows the cut used. 
Right: the $J/\psi\phi$ invariant mass distribution after Belle's final selection. 
Data are represented as the  open histogram. The total fit of the data and background component
are the solid and dashed curves respectively.  The shaded histogram represents the background  
normalized from the $J/\psi$ and $\phi$ mass sidebands.
\label{f12belletwophotonf1}}
\end{figure}

The Belle collaboration also extended their search in this channel to  two-photon process 
using the same dataset~\cite{belleevidence}. 
The  $\mu^+\mu^-K^+K^-$ and $e^+e^-K^+K^-$ events with the 
four tracks pointing back to the interaction point are selected. 
Figure~\ref{f12belletwophotonf1} (a) shows the magnitude of the sum 
of transverse momentum vectors making up the 
four-body system, with respect to the beam position, 
and a cluster of events around the zero value is observed. 
Those events with the magnitude of the sum 
of transverse momentum vectors to be less than 0.2 GeV are selected to 
be candidates from the two-photon process.
Figure~\ref{f12belletwophotonf1} (b) shows the 
invariant mass of the combined $\mu^+\mu^-K^+K^-$ and $e^+e^-K^+K^-$ 
system in the $J/\psi\phi$ signal region and its sideband regions  for the selected events. 
They found no evidence of the $Y(4140)$ 
in two-photon process, however, they did find evidence for a new structure 
 with a mass of $4350.6^{+4.6}_{-5.1}(\mathrm{stat})\pm0.7(\mathrm{syst})$ MeV, 
  and a width of  $13^{+18}_{-9}(\mathrm{stat})\pm4(\mathrm{syst})$ MeV 
called $X(4350)$ with a significance of 3.2$\sigma$. 
This structure cannot have  $J=1$, being visible in the two-photon process and must have a positive 
charge parity.
If confirmed, it is another good candidate 
for an exotic meson.

\subsection{Structures in the $J/\psi\phi$ Mass Spectrum: LHCb}
\indent

\begin{figure}[pb]
{\psfig{file=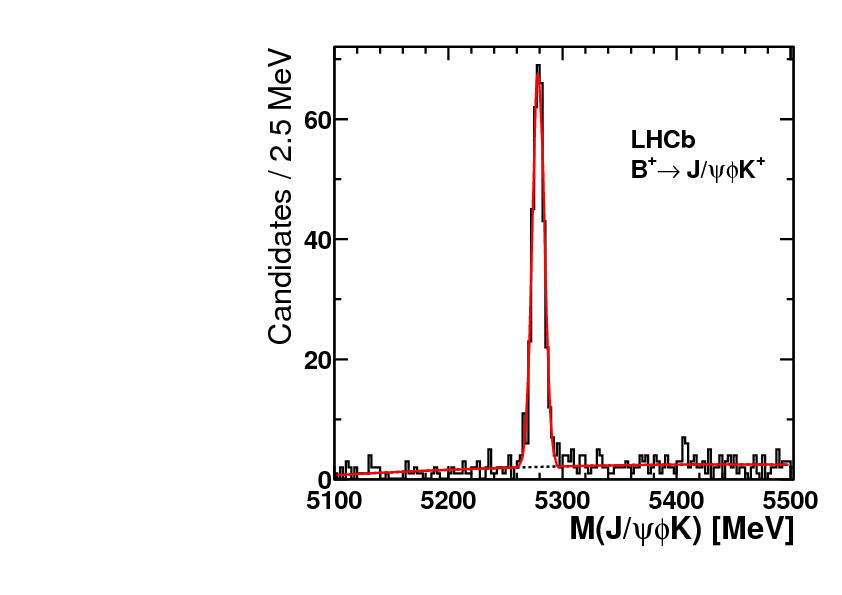,width=0.495\textwidth}}
{\psfig{file=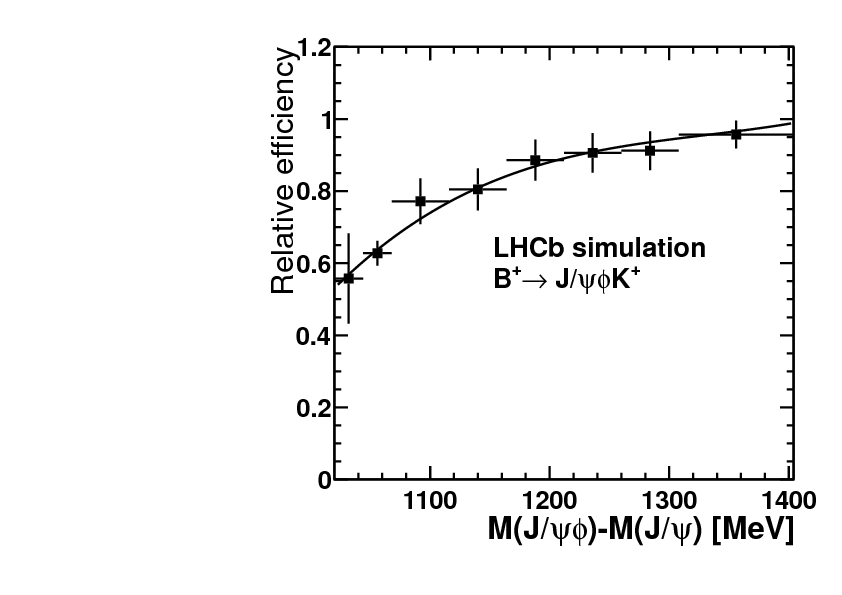,width=0.495\textwidth}}
\centerline{\psfig{file=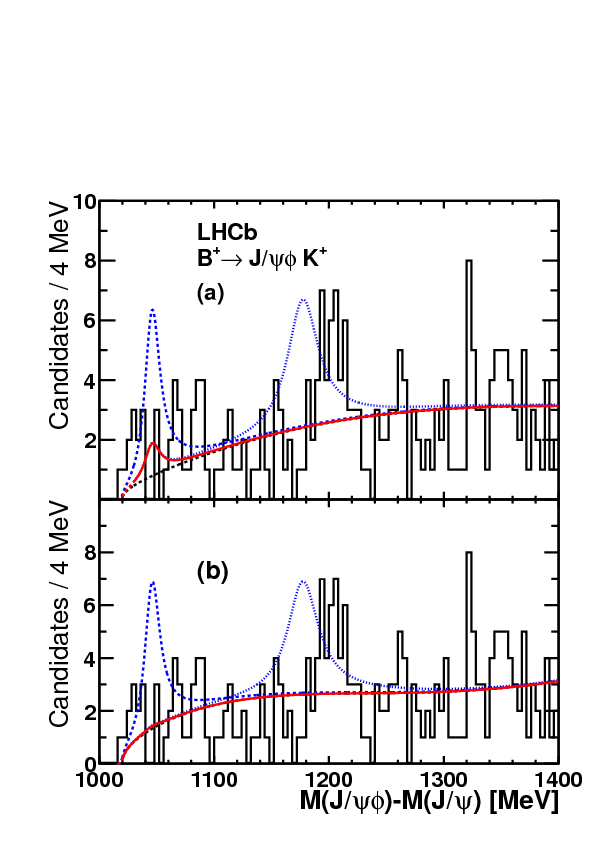,width=0.5\textwidth}}
\vspace*{8pt}
\caption{
Top left: the mass distribution of $J/\psi\phi K^+$ at LHCb.
A fit to the data with a Gaussian signal function and flat
background function is shown as the solid red line.  
Top right: the relative efficiency for $J/\psi\phi$ as a function of the mass difference, $\Delta M$, 
between $\mu^+\mu^-K^+K^-$ and $\mu^+\mu^-$, for LHCb.
Bottom: The $\Delta M$ distribution in the $B$ mass window (a) using three--body phase space background 
shape (b) using three--body phase space multiplied by a quadratic function at LHCb. 
The total fit  is represented as the solid red  curve   
and the dash-dotted blue curve 
shows the expected signals estimated from CDF results.
\label{f14lhcbf1}}
\end{figure}

In 2011, the LHCb collaboration searched for the $Y(4140)$  through the 
exclusive $B^+ \rightarrow J/\psi \phi K^+$ channel as CDF, but
using 370 pb$^{-1}$ of data collected from 7~TeV $pp$ collisions~\cite{lhcbjpsiphik}.
At LHCb, they either require a $J/\psi \rightarrow \mu^+\mu^-$ 
with $p_T>1.5$ GeV or a muon-track pair with significant IP to confirm trigger requirements.
In order to reduce beam related background, LHCb requires 
a minimum transverse momentum of 250 MeV  for reconstructed kaon tracks. 
The forward position makes  LHCb's relative detection  efficiency  about 40\%  
lower near the $J/\psi\phi$ threshold region where kaon tracks normally have low 
transverse momentum with lower efficiency.
LHCb then combined other selection  information into a likelihood 
ratio to select the $B$ signal events.  
They reconstructed 346$\pm$20 
$B^+$ events with a mass resolution of 5.2 MeV and negligible non-B background. 
The $J/\psi\phi K^+$ invariant mass distribution, 
the relative efficiency  as a function of $\Delta M$,  
and the $\Delta M$ distribution 
in the $B$ mass window are shown in Fig.~\ref{f14lhcbf1}.
The LHCb experiment has about three times   
CDF's statistics, however, the sensitivity is  
reduced due to the efficiency drop (by 40\%) 
near the $J/\psi\phi$ threshold.
LHCb found no evidence  for the $Y(4140)$ and they estimate that they are  
in 2.4$\sigma$ disagreement with 
CDF's report. An upper limit on its branching fraction relative to 
the exclusive $B^+\rightarrow J/\psi \phi K^+$ is quoted to be $<$ 0.07 at 90\% 
confidence level using the $Y(4140)$ parameters from CDF.
Furthermore, they also set an upper limit for the second peak on the branching 
fraction relative to the exclusive $B^+\rightarrow J/\psi \phi K^+$ to be $<$ 0.08 at 90\% 
confidence level using the $Y(4274)$ parameters from CDF.
The results from the LHCb and CDF stand in 
contention~\footnote{A recent unofficial result from LHCb using 1.0 fb$^{-1}$ seems 
supportive of  the existence of the $Y(4140)$ structure, although no fit parameters were 
quoted to enable a quantitative comparisons~\cite{lhcbunofficial}.}.

\section{A New Kind of Spectroscopy in the $J/\psi\phi$ System?} 
\indent

\subsection{The Status of the $J/\psi\phi$ Structures}

\begin{figure}[pb]
{\psfig{file=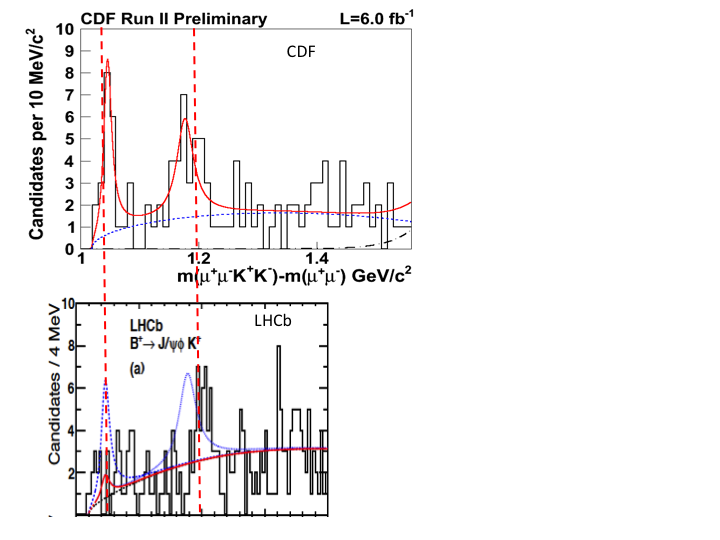,width=1.5\textwidth}}
\vspace*{8pt}
\caption{
A direct comparison of $J/\psi\phi$ mass or $\Delta M$ distributions among CDF,  LHCb.
\label{f15comparison}}
\end{figure}

\begin{figure}[pb]
{\psfig{file=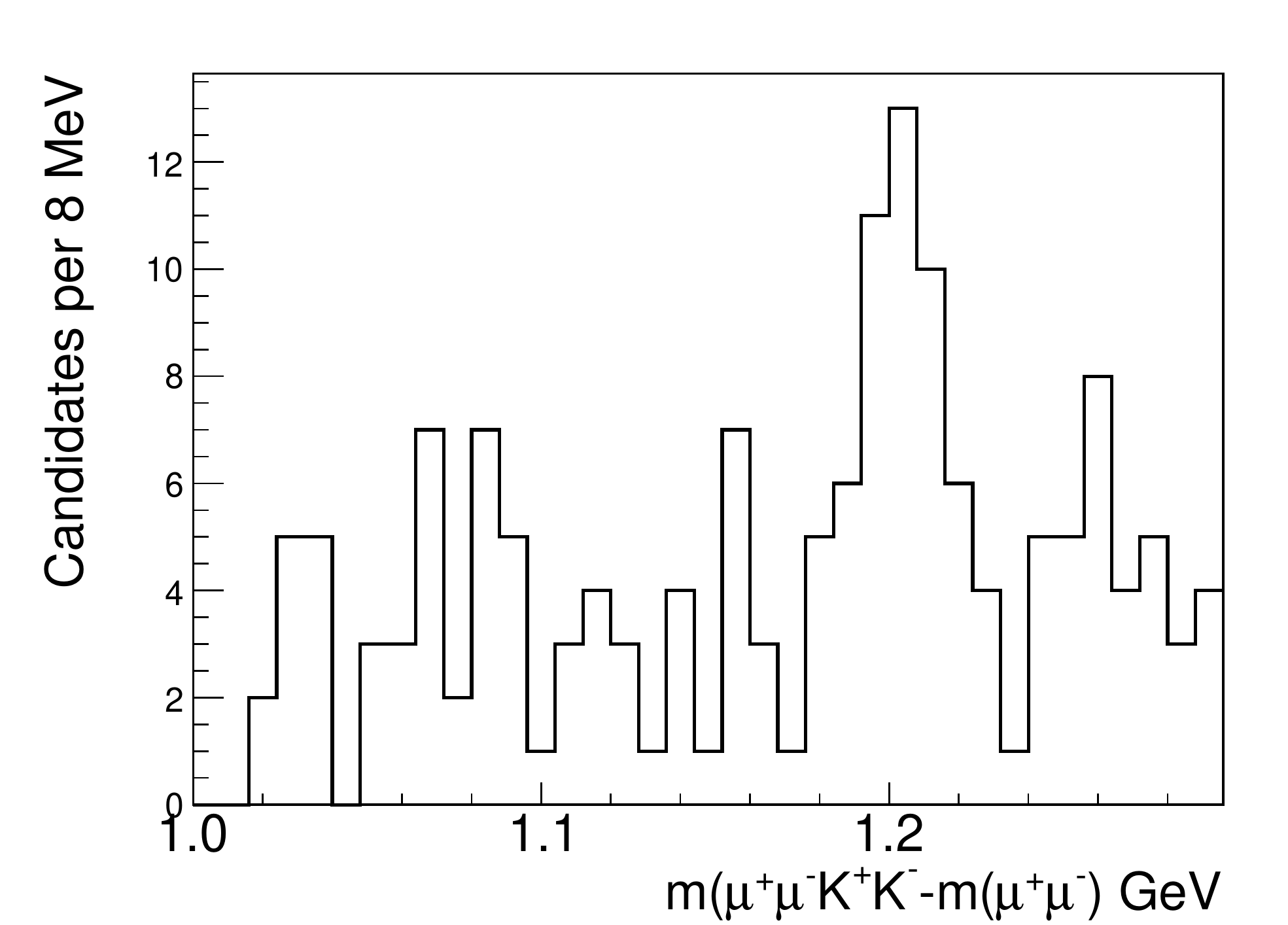,angle=90,width=0.48\textwidth}}
{\psfig{file=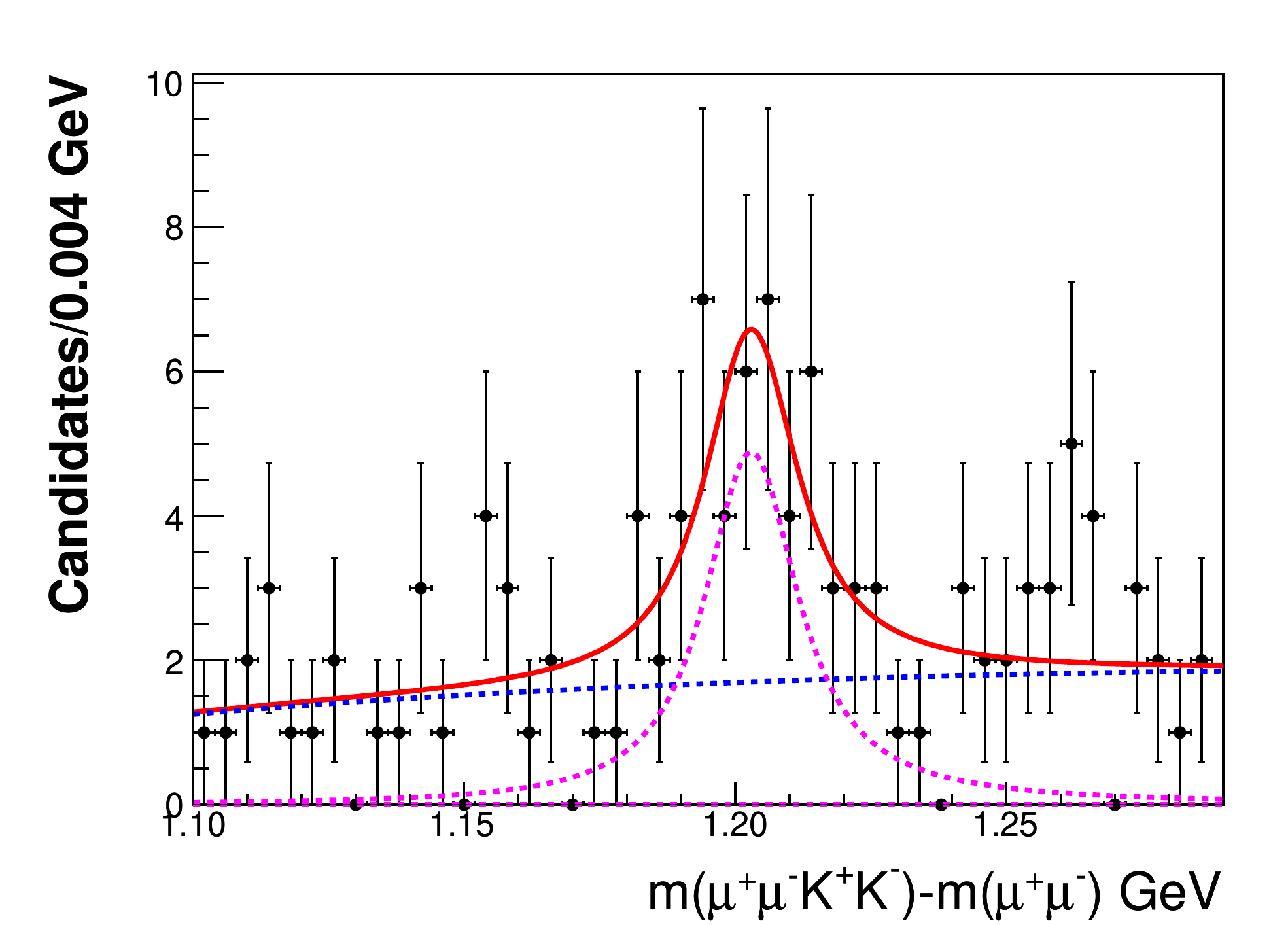,angle=90,width=0.48\textwidth}}
\vspace*{8pt}
\caption{
Left: The $\Delta M$ distribution of LHCb data in the neighborhood of CDF's second structure
obtained by rebinning their data into 8 MeV bins.
Right: A fit to the apparent excess in LHCb data using a Breit-Wigner function  plus 
a three-body phase space background shape.
}
\label{f16lhcbrebin}
\end{figure}

There are at least three possible structures that have been experimentally claimed 
in the $J/\psi\phi$ mass spectrum, 
none of them are expected from the conventional  charmonium  states, and are thus
good candidates for  exotic mesons.
Even though there has been contention between LHCb results and those of CDF, 
There are obvious activities around the $Y(4140)$ and $Y(4274)$ regions in the 
$J/\psi\phi$ mass spectrum. A paper from CMS on $J/\psi\phi$ structures based on the largest 
$B\rightarrow J/\psi\phi K^+$ sample in the world to date  is expected to be 
submitted to {\it Physics Letter B} soon. 
LHCb also has ten times of more data 
to be analyzed, and thus we expect greater clarity in the near future.

To further examine the status of the $J/\psi\phi$ structures  in exclusive $B$ decays,  
a display directly comparing the 
$J/\psi\phi$ or $\Delta M=m(\mu^+\mu^- K^+ K^-)-m(\mu^+\mu^-)$ 
mass spectra from two of the    
experiments is shown in Fig.~\ref{f15comparison}. 
Consider first 
the most significant structure originally reported by CDF, the $Y(4140)$ 
in 2009. The statistics for CDF's report is low and it is  difficult to 
determine  the parameters of this structure, 
and  confirmation from independent experiments was vital.
The absence of any excesses in LHCb's report in 2011  for CDF's 
 $Y(4140)$ is quite striking. 
The discrepancy can be understood quantitatively from the expected signal yields:
the LHCb fitted yield using CDF parameters is 6.9$\pm$4.9 candidates 
compared to an expected yield 
scaled from CDF's of 35$\pm$9$\pm$6 candidates, 
a 2.4$\sigma$ disagreement as estimated by LHCb~\cite{lhcbjpsiphik}.
The ambiguous situation over the  status of the $Y(4140)$ presented 
the usual dilemma of whether statistical fluctuations have confused 
the picture or unappreciated instrumental effects had struck?
And if the latter, for which experiment?

On the other hand, there is in fact considerable 
consistency among experiments if one allows for the large uncertainty 
in ascertaining the properties of broad low-statistics structures 
above the near threshold region.
One readily notices
in Fig.~\ref{f15comparison} that there 
actually seems to be an excess around $\Delta M=1.2$ GeV, 
or $m(J/\psi\phi)=4.3$ GeV, among the two mass spectra  
even though they are of low significance and 
are not  exactly aligned to the same position.

An excess seems quite pronounced at higher mass in the LHCb spectrum, 
which becomes visually  more 
striking when their data is rebinned to 8 MeV, as shown in Fig.~\ref{f16lhcbrebin} (lef).
An estimation from my simple fit of this excess, shown in Fig.~\ref{f16lhcbrebin} (right),
gives the rough parameters of a possible 
second structure as: mass $4.300\pm 0.004$ GeV,  
width $20\pm10$ MeV, yield  $36 \pm 12$, 
and a local significance of about 3$\sigma$ 
(in each case with statistical uncertainties only).  
The simple model assumed here did not have the efficiency correction,
and the mass resolution was assumed to be 2 MeV.  The excess was 
represented by a Breit-Wigner convoluted with Gaussian resolution function, and 
the background was a three-body phase space shape.
We are left with the intriguing prospect that there probably is 
evidence for this second $J/\psi\phi$  structure  in LHCb data---although 
it is not in very good accord with CDF's parameters.
The latter fact helps explain  LHCb's negative report of a signal:
because they {\it imposed} CDF parameters in their search the apparent yield 
was compatible with background.
However, given the difficulties of extracting mass parameters for broad weak signals 
the poor consistency between CDF and LHCb may not be so surprising.

The width and relative BF for the second structure are consistent among the two experiments.
However,  the mass  is about 2.4$\sigma$ 
inconsistent between LHCb and CDF's result assuming negligible systematic 
uncertainty from LHCb experiment.

Overall, there is obvious activity going on around 
$\Delta m=1.2$ GeV or $m(J/\psi\phi)=$ 4.3 GeV.
However, the locations from various 
experiments are not in very good agreement, but yet they are not entirely inconsistent. 
More data from the LHC should fully resolve the situation.

\subsection{Possible Interpretations and Outlook}

Various explanations have been proposed for these structures in
the $J/\psi\phi$ spectrum, but their nature is still not understood. Further joint
experimental and theoretical effort is needed to completely understand their 
origin. I will summarize the possible interpretations based on what 
is currently known including conventional charmonium, molecular, hybrid, non-resonance 
effects.
Recently there was also a hadro-charmonium proposal which is a QCD analog of 
the van derWaals force~\cite{hadrocharm,hadrocharmextend}.
Hadro-charmonium is a new type of bound state formed
by binding a relatively compact charmonium state
inside an excited state of light hadronic matter.
The charmonium 
state behaves like a compact sub-object inside the new state such 
that the loosely bound  states decay into this charmonium and light mesons. 
However, there is no detailed calculations for $J/\psi\phi$ structures so far and 
I therefore pass over it in the following discussion.

{\bf ${\bf Y(4140)}$ options}. 
{\bf  Conventional charmonium}:
The $Y(4140)$ mass is well beyond the open charm threshold, and thus it would be
expected to decay dominantly into open charm pairs with a large width and to
decay into $J/\psi\phi$ with a tiny branching fraction.
For instance, the upper limit on the 
branching fraction of $Y(4140)$ as  the second radial excitation of the $P$-wave 
charmoniums $\chi^{\prime}_{cJ} (J=0,1)$ decaying into hidden charm 
is calculated to be in the order of $10^{-4}$ to $10^{-3}$~\cite{Liu:2009iw}.
This implies that the $J/\psi\phi$ is not likely to be a discovery channel 
for these conventional $\chi$-states, and correspondingly, that the 
$Y(4140)$ is unlikely to be a conventional charmonium state.
{\bf  $\bf  { D^{*+}_sD^{*-}_s}$ molecule}:
Due to the  similarity 
between $Y(4140)$ and $Y(3940)$~\cite{belley3940,babary3940}, 
such as their decay channels and near threshold masses, 
the $D_s^{*+}D_s^{*-}$ 
and $D^{*}\bar{D}^{*}$ molecular state explanations to $Y(4140)$ and $Y(3940)$ 
were proposed in Ref.~\cite{Liu:2009ei}, respectively.
The $D^{*+}_sD^{*-}_s$ threshold is about 80 MeV above the $Y(4140)$ mass, 
and thus the system would have a binding energy of ~100 MeV. The $D^{*+}_sD^{*-}_s$ molecule 
is bound together through attraction provided by meson exchange. 
It decays into two-body  hidden-charm final states via the re-scattering of $D^{*+}_sD^{*-}_s$ 
and into two-body open charm states via the exchange of light mesons,
and it does so with roughly equal 
probability~\cite{Liu:2009ei,Mahajan:2009pj,Branz:2009yt,Zhang:2009vs,Albuquerque:2009ak,Ding:2009vd}. 
Each sub-system in the molecule could  decay into $DK$ pair,  but it is 
not kinematically  allowed ({\it i.e.} $DK$ mass is above $D^{*}_s$), thus the $D^{*+}_sD^{*-}_s$ molecule 
has narrow  natural width. 
To further test the molecular hypothesis of $Y(4140)$, 
the experimental measurement of the photon spectrum of 
$Y(4140)\rightarrow {D}_s^{\ast+} D_s^- \gamma, D_{s}^+D_{s}^{*-}\gamma$ 
are suggested. The  $D^{*+}_sD^{*-}_s$ molecule seems a viable interpretation of $Y(4140)$, 
but no concrete conclusion can be made. Furthermore, the two-photon partial width measured 
by Belle experiment~\cite{belleevidence} disfavor the interpretation of  $D^{*+}_sD^{*-}_s$ molecule 
with $J^{PC}$= $0^{++}$ or  $2^{++}$ for $Y(4140)$.
{\bf  ${\bf c\bar{c}q\bar{q}}$ tetra-quark state}:
The $J/\psi\phi$ decay channel of $Y(4140)$ contains $c\bar{c}$ and 
$s\bar{s}$ components, thus it was also considered as a 
$c\bar{c}s\bar{s}$ tetra-quark state without sub-systems inside through rearrangement of the constitute 
quarks with a width of about 100 MeV, and about 
equal probabilities for both hidden and open-charm 
two-body decays~\cite{Stancu:2009ka}.
The observed width does not seems  support this proposal.
{\bf  Charmonium hybrid state}:
Since the $Y(4140)$ mass 
is in the range of 4.1-4.4 GeV as predicted by various models
for charmonium hybrid states,  
it  is also considered  as a charmonium hybrid 
candidate~\cite{Mahajan:2009pj}.  
The width of a hybrid is expected to be broad, however
if it has exotic $J^{PC}$ of $1^{-+}$ and its mass  lies below
the threshold of $D^{**}D$, then its width can be narrow~\cite{hybridselection}.
This model predicts $D^*\bar{D}$ an an important decay.
{\bf  Non-resonance effect}:
Besides exotic resonance proposals, there are  non-resonance proposals. 
For instance,  a kinematical  effect due to the opening of the $J/\psi \phi$ 
channel, with the creation of a $s\bar{s}$ pair and 
further forming a $\phi$ resonance surrounding the $c\bar{c}$ pair, 
the cross section rises and then falls rapidly to produce a
peak-like structure, but this has nothing to do with being
a resonance~\cite{vanBeveren:2009dc}.

{ \bf  ${\bf Y(4274)}$ options}.
{\bf  Conventional charmonium}:
The $Y(4274)$ mass is even further beyond the open charm threshold compared 
to $Y(4140)$, and thus it is also very unlikely to be conventional charmonium.
{\bf   ${\bf \bar{D}_sD^0_{s}(2317)}$  molecule}:
Since the mass of $Y(4274)$ is close to  $\bar{D}_sD^0_{s}(2317)$, 
like the proximity between  $Y(4140)$ mass to the $D_s^{*+}D_s^{*-}$ threshold,  this structure is 
also proposed as a possible $\bar{D}_sD^0_{s}(2317)$ 
molecular charmonium state~\cite{Shen:2010ky,Liu:2010hf,y4274molecule,y4274notmolecule}. 
This proposal predicts other channels similar to $Y(4140)$ decays, 
such as $Y(4274)\rightarrow D^+_s D^{*-}_s \gamma$ 
and $Y(4274)\rightarrow D^+_s D^{-}_s\pi^0$ etc, which should be searched for.
{\bf  Charmonium hybrid state}: 
Another possibility is that this structure can be a hybrid charmonium. 
A recent Lattice QCD calculation of  hybrid charmonium  with an exotic $J^{PC}$
of $1^{-+}$ yields a mass around the observed mass of this 
structure~\cite{Liu:2012ze}.

{\bf  ${\bf X(4350)}$ options}.
{\bf  Conventional charmonium}:
The $X(4350)\to J/\psi \phi$, similar to $X(3915)\to J/\psi \omega$, 
was discovered in the two-photon process, 
which is an abundant source of charmonium production via $\gamma\gamma$ fusion.
The $J^{PC}$ possibilities of $1^{-+}$, $1^{++}$ 
and $3^{-+}$ are excluded due to Yang's Theorem.  
Under the $P$-wave charmonium assignment to $X(3915)$ and $X(4350)$, the $J^{PC}$ 
quantum numbers of these two states must be $0^{++}$ and $2^{++}$ 
respectively, which shows that $X(3915)$ and $X(4350)$ can be as the second and third 
radial excitations of $\chi_{c0}$ and $\chi_{c2}$~\cite{Liu:2009fe}, respectively. 
And these authors found that the calculated mass spectrum and two-body $OZI$-allowed 
strong decay behavior are consistent with experimental results.
{\bf   ${\bf D^*_s \bar{D}^*_s}$  molecule}: 
The $X(4350)$ is also proposed as 
a $D^*_s \bar{D}^*_{s0}$ molecular state~\cite{x4350dsds} as well as 
charmonium-virtual molecular ($c\bar{c}$-$D^*_s \bar{D}^*_{s}$) 
mixing state~\cite{Wang:2009wk}.
As this is only observed in two-photon process so far, 
the Belle~II program would provide the large increase in yield that will
make further investigations possible before it appears in other process 
at the LHC.

Finally, there are also a number of interesting phenomenological observations 
to be made. 
First, there are similar near threshold excesses which have already 
been observed in light vector mesons through 
radiative $J/\psi$ or exclusive $B$  decays:  
the near threshold enhancements for  
two isospin-0 vectors $\phi\phi$, $\omega\omega$, 
and $\omega\phi$~\cite{phiphiref1,phiphiref2,phiphiref3,phiphiref4,omegaomegaref,omegaphiref1,omegaphiref2}; and 
the near threshold enhancement for two isospin-1 
vectors $\rho\rho$~\cite{rhorhoref1,rhorhoref2}.  
However, there has been no clear structure observed that decays into 
two  vectors that one of them has isospin-1 and the other 
has isospin-0--$\phi\rho$ or $\omega \rho$~\cite{rhovref}. 
These  near threshold enhancements for  
two light isospin-0 vectors could possibly be connected to 
the $Y(3940)\rightarrow J/\psi\omega$ 
and $Y(4140)\rightarrow J/\psi \phi$ since both 
$J/\psi$ and $\phi$ have isospin-0 and are very close to the two vector 
threshold.
Second, it is striking that there are at least two structures observed 
in the same $J/\psi\phi$ spectrum from the same exclusive $B$ decays. 
The two structures do not need to arise from the same mechanism, but 
if they are, this could be a pointer to  a new kind of spectroscopy. 
It will also be interesting to search for  vector-vector 
structures composed entirely of c and b quarks near threshold 
because they may offer simpler
systems to model theoretically.

\section{Summary}
\indent

The discovery of the $X(3872)$ in 2003 initiated a new chapter 
in the study of exotic states. The evidence found in the $J/\psi\phi$ 
mass spectrum from exclusive $B$ decays and two-photon process 
provide several exotic charmonium candidates: they all have a  mass well 
beyond the open-charm pair threshold, a relatively narrow width, positive C-parity, 
and do not fit into the conventional charmonium picture.  
All of these reported structures need further studies.  
Various possible interpretations such as molecule, tetra-quark, charmed hybrid, 
nuclei-like structures have been proposed but none of them is established.  
The $pp$ run at the LHC is now over until 2015, but CMS has roughly four 
times more data on tape to be analyzed and LHCb is expected to have about 
ten times the data to be analyzed. ATLAS, on the other hand, has not yet been heard from but
they can also contribute  to the current studies. 
The properties of these structures,   such as mass, 
width, decay channels and quantum numbers, can be studied 
with increased statistics. At the LHC one can also search in other heavy quarkonium decay channels 
for vector-vector analogs of the $J/\psi \phi$ structures.
With joint experimental and theoretical effort, hopefully we
will obtain a clear grand picture and eventually understand the nature
of these queer structures---which may indeed lead us to
a new kind of spectroscopy.

\section*{Acknowledgments}

I would like to thank  
Gerry Bauer, Stanley Brodsky, Joel Butler, Tiziano Camporesi, Kai-feng Chen, 
Vincenzo Chiochia, Ray Culberson, Su Dong, Tommaso Dorigo, William Dunwoodie, Estia Eichten, 
Bryan Fulsom, Tim Gershon, Christopher Hill (OSU), Pat Lukens, Namit Mahajan, 
Gautier Hamel de Monchenault, Jane Nachtman, 
Stephen Olsen, Fabrizio Palla, Frank Porter, Weiguo Li, Xiang Liu, 
Xinchou Lou, Rob Roser, Jonathan Rosner, Randy Ruchti, Paraskevas Sphicas, 
Christopher Thomas,  Jim Russ, Joao Varela, Mikhail Voloshin, 
Hermine Woehri, E. Asli Yetkin, Shilin Zhu as well as CMS quarkonium group  
for useful discussions.
However, all opinions and comments expressed, and any error committed, 
are solely responsibility of the author himself.

\end{document}